# Observation of a nanophase segregation in LiCl aqueous solutions from Transient Grating Experiments

**L. E. Bove[1], C. Dreyfus[1], R. Torre[2], and R. M. Pick[1a]**

1 *IMPMC, Université P. et M. Curie et CNRS-UMR 7590, Paris (France)*

2 *LENS and Dip. di Fisica, Università di Firenze (Italy)*

## Abstract

Transient Grating experiments performed on supercooled LiCl, $RH_2O$ solutions with R>6 reveal the existence of a strong, short time, extra signal which superposes to the normal signal observed for the R=6 solution and other glass forming systems. This extra signal shows up below 190 K, its shape and the associated timescale depend only on temperature, while its intensity increases with R. We show that the origin of this signal is a phase separation between clusters with a low solute concentration and the remaining, more concentrated, solution. Our analysis demonstrates that these clusters have a nanometer size and a composition which are rather temperature independent, while increasing R simply increases the number of these clusters.

[a] a Corresponding author e mail :robert.pick@courriel.upmc.fr

**A Introduction**

At normal pressure (1 bar), liquid water crystalizes at 273 K into the hexagonal ice $I_h$ with a mass density equal to $0.92g/cm^3$, while the mass density is equal to $1g/cm^3$ for the liquid. In the $I_h$ structure, each water molecule is tetragonally coordinated to its four nearest neighbors. Liquid water is also usually classified as a tetrahedral liquid because its coordination number, defined as the area under the first peak of the oxygen–oxygen radial distribution function, $g_{OO}(r)$, actually increases slightly upon melting of hexagonal ice, even if the four directional hydrogen bonds of the solid are distorted by thermal fluctuations to permit less ideal and more compact hydrogen-bonding arrangements in the liquid. Also, cubic ice, which can be formed for instance by the condensation of water in its vapor phase at temperatures around 190 K, has the same n=4 coordination number as ice $I_h$.

By very slow cooling, water can be supercooled below 273 K, but it inevitably crystalizes into ice above 240K. One needs to hyper quench small water droplets, at cooling rate faster than $10^6 K/s$, down to liquid nitrogen temperature to obtain an amorphous solid. Through the preceding technique, one forms a glass, called HGW (hyperquenched glassy water), which is structurally equivalent to LDA (Low Density Amorphous) water [1] the phase obtained by transformation under temperature annealing at ambient pressure of the HDA (High Density Amorphous) phase. LDA also has approximately the same low density as ice $I_h$, with the same coordination number, and thus local environment, the disorder appearing only in the absence of long range order, while HDA, which is obtained by, for instance, compressing ice $I_h$ at or below 125 K up to a few hundreds of MPa has a structure which is more similar to high density liquid water (HDL) [2]. The principal feature of its structure, which can slightly vary following different production paths, is a puckering (collapse) of the more distant molecules, increasing the number of interstitial, non-hydrogen bonded, water molecules in the first coordination shell. Releasing the pressure this HDA water remains metastable at ambient pressure at temperatures below 140 K with a mass density equal to $1.15g/cm^3$ [3].

The two preceding amorphous states have been obtained by irreversible techniques, and the impossibility of obtaining glassy water by a continuous and slow cooling technique is at the origin of the notion of a "no-man's land" [4] in the phase diagram of water. This has also led to seek for other strategies, as confinement in nano-pores [5] or addition of non-crystallizable substances [6], for forcing water in a deep undercooled state, below ice homogeneous nucleation temperature. One of the less invasive, and mostly used, ways to deeply undercool water is to add small size ions. In particular, since the pioneering work ok Vuillard and Kessis [7], it is known that LiCl, RH$_2$O can be supercooled, in the vicinity of its eutectic point (R=7.06 and T=198 K), down to its glass transition, which is around 140 K. The study of these LiCl, RH$_2$O solutions has been very fashionable since the first measurements of Angell and Sare [8]. In particular, Elarby-Aouizerat *et al* [9 (a)] showed that the R domain where the glass could be produced with a very slow cooling rate was in fact located, not around R=7 but rather around R=6. This points out, for this solution, to a possible competition between a chemical effect (fixed number of water molecules for one Li$^+$ Cl$^-$) and the usual thermodynamic one (equality of thermodynamic potentials at the eutectic concentration and temperature, both quantities varying with pressure). Another aspect of these solutions is worth mentioning. Between R=∞ and R=7.06, there is a temperature domain where the liquidus is in equilibrium with a solidus, i. e., in principle, a water-LiCl solid solution; in the present case, this solidus is practically a pure ice phase, the LiCl salt remaining totally dissolved in the liquid phase. In other words, the salt is completely expelled from the ice lattice. As the volumes of the Cl$^-$ ion and of a water molecule are practically the same, this can be interpreted as the absence of a site which can hold a Li$^+$ cation in the open structure of ice I$_h$ or of LDA water. Conversely, it has been found that, in the R=6 glass, the Li$^+$ cation is trapped in the octahedral environment of six water molecules [10]. This is true, in particular, at high pressure where exists an "ordered" structure, similar to ice VII, in which Cl$^-$ anions randomly replace water molecules while the corresponding small Li$^+$ cations are in interstitial positions [11]. Because the Br$^-$ anion is only slightly larger (by some 8%) than the Cl$^-$ one, the same could also be the case for the LiBr, 6H$_2$O solution. Nearly octahedral coordination seems to be favored also for bigger cations as K and Na in solution [12].

The preceding considerations suggest that the study of the LiCl, 6H$_2$O supercooled liquid could give hints on the deeply undercooled state in water, the replacement of one out of seven water molecules by a Li$^+$Cl$^-$ "doublet" preventing the change of structure from liquid water to ice I$_h$ but not its ability to supercool. Furthermore, as the electrostrictive effect provided by the ions on water molecules changes their structural organization in an equivalent way to the effects of an applied external pressure [11,13] we can in such a way probe the HDL-HDA boundary, otherwise inaccessible.

In this context we have undertaken a series of Heterodyne Detected Transient Grating (TG) experiments at the LENS Laboratory (I) on LiCl and LiBr-RH$_2$O solutions, with different values of R, equal to or larger than 6. The goal was to explore the R region where there is a competition between the two mechanisms which lead to the formation of a glass through a slow cooling rate The experiments have been performed from room temperature down to 160 K, a temperature 20 K above the glass transition T$_g$=140 K [14, 15]

 The LiCl, 6H$_2$O and LiBr, 6H$_2$O, R=6, TG experiments, complemented by Brillouin and ultrasonic experiments, have been carefully analyzed in the whole temperature region and lead to very similar results for the two anions [16]; the longitudinal phonons generated by the pump laser (see Section B) are coupled to two relaxation processes: a usual α process whose relaxation time, τ$_α$, diverges in a Vogel-Fulcher like manner, and an additional β relaxation process which becomes apparent only below 210K and whose relaxation time τ$_β$ exhibits an Arrhenius-like behavior with a value not exceeding 60 ps at T$_g$ [16]. Conversely, the R>6 solutions had a very different low temperature behavior. While the bromine solutions had similar thermal behaviors whatever R, an extra, short times signal appeared in the chlorine case. Analyzed as a signal which superposes to the "normal" R=6 one, the intensity of this "extra signal" turned out to increase with R while its duration and shape depended only on temperature. This extra signal had ben already noticed, more than 15 years ago, by Nelson *and al.* [17], for R=7.7 but the poor resolution of the TG sets up available at that time did not allow for its proper analysis. The purpose of this paper is to clearly characterize this signal and to explain its origin. In order to achieve this goal, the present paper is organized as follows.

Section B recalls the two origins of the signal detected in a HD-TG experiment. Section C describes the experimental results obtained at various temperatures, R values, and wavevectors. The next section, Section D, proposes an explanation for the origin of the extra signal and Section E develops the corresponding mathematical formulation. In Section F, we analyze our experiments with the help of these formulae and discuss the physical meaning of the obtained parameters. We also compare our findings with previous experimental and/or computational results. The final Section G summarizes our results and proposes some different experiments to further study those clusters.

**B Transient Grating Experiments: a brief summary**

Let us briefly summarize the two origins of the signal recorded in a usual HD-TG experiment [18], giving, simultaneously, the characteristic times that are involved in the LiCl, 6H$_2$O case. The TG technique is a pump-probe, time-resolved, experiment in which a very high intensity, nearly instantaneous, pulse (duration of the order of $10^{-1}$ ns), with a frequency in the near infrared, is divided into two coherent beams (the pumps) of equal amplitudes which propagate in quasi parallel directions. They interfere in the liquid and generate, *in fine*, two density gratings of wavevector $\vec{q}$. The time evolution of their sum is monitored by the amplitude, $S(t)$, of a probe beam diffracted by these gratings and heterogenely detected, this amplitude being proportional to the change of the index of refraction, $\delta n$, of the liquid [18,19], thus to its local density change, $\delta \rho$.

The first grating has an electrostrictive origin. The modulated electrostrictive force, created by the interaction of the interference electric field of the two pumps with the liquid, generates two longitudinal phonons. Those two phonons have equal amplitude, wave vectors $\vec{q}$ and $-\vec{q}$, and a period of a few ns. Their interference creates a time dependent density grating with the same spatial and temporal periods, called the ISBS (Instantaneous Stimulated Brillouin Scattering) grating. Its duration is governed by the phonons life time and, for the temperatures and the water solutions probed here, this lifetime extends from a few ns (190 K) to $\approx$ 80 ns (172 K).

The second grating, which will also be the source of the extra signal we shall discuss later, is of thermal origin and exists only for molecular liquids. The total electric field generated by the interference of the pumps has a spatially modulated amplitude, and is weakly absorbed by the overtones of the intramolecular vibrations. These overtones having a very short lifetime, the absorbed energy is, instantaneously and locally, transferred to the heat bath, creating a temperature grating with the same wavevector $\vec{q}$. This thermal grating induces a pressure grating which, in turn, generates phonons identical to the first ones. Also, the local density of the liquid tends to adjust to this local temperature, generating a static density grating. The amplitude of the latter will eventually decrease due to heat diffusion from the hottest parts of the grating to the coldest ones. The heat diffusion time, $\tau_h$, is proportional to $q^{-2}$ and, in the present experiment, decreases from $\approx 2 \; 10^4$ ns for q=0.63 mm$^{-1}$ (most of the experiments reported here) to $\approx 2 \; 10^3$ ns for the largest wavevector used (q=1.76 mm$^{-1}$). We checked in ref. [16] that $\tau_h$ does not depend on the temperature in the temperature range studied. The diffraction due to this thermal grating is called Instantaneous Stimulated Thermal Scattering (ISTS) and we shall show, in the next Sections, that the extra signal has the same thermal origin and is due to the fact that, in LiCl, RH$_2$O with R>6, the thermal (thus density) grating has also a short time evolution not included in the preceding elementary description.

## C Experimental Results

TG experiments on LiCl, RH$_2$O were performed at a variety of temperatures, R values and wavevectors. A large number of experiments were concerned with the R=6 solutions at q=0.63 $\mu$m$^{-1}$ at various temperatures between 300 K and 166 K. Their analysis has been briefly summarized in the Introduction and is fully reported in [16(a)]. Furthermore, some experiments were performed with the same q value on solutions with R=6.45, 6.66 and 7.14. They revealed at temperatures equal to and lower than 190 K, the extra signal we discuss in the present paper. Finally a series of experiments were performed at 181 K with the R=6.66 solution as a function of the wavevector, namely for q=1, 1.38, and 1.76 $\mu$m$^{-1}$. Table I lists all the experiments which could be analysed in the framework of the present study.

Fig. 1a shows the result of those experiments (amplitude of the diffracted beam measured as a function of the time difference between the pumps and the probe pulses, for T=184 K and q=0.63 $\mu$m$^{-1}$, for the four R values mentioned above. The four signals exhibit exactly the same long time behaviour (t>300 ns), characteristic of the final decay of the ISTS signal. Each of them also exhibits, at very short times, an oscillatory behaviour with a period of the order of 3.3 ns, typical of the interference between the two longitudinal phonons generated by the pumps. A close inspection of the four signals shows that this period slightly increases with R, indicating an apparent sound velocity increase by a few per-cents in the R=6-7.14 interval at that temperature.

Fig. 1a also shows that the amplitude of the signals increases with R-6 at times shorter than 300 ns. Nevertheless, the shape of the extra signals is quite independent of R: their long time behaviour (20 ns<t<300 ns) is exactly the same for R=6.6 and R=7.14 (see Fig. 1b), while this shape slightly differs for R=6.45. This small difference is due to the weak intensity of the R=6.45 extra signal, which makes a difference signal very sensitive to a small temperature difference between the two solutions. This is true for all temperatures and explains why we did not analyse these R=6.45 extra signals. Finally, the differences at very short times (t<20 ns) between signals with different R result from the R dependence of the apparent sound velocity. Let us stress that the R=6 signal and the extra signals must have the same physical origin: varying the pump intensity, the relative shape of the two signals does not change.

Figs. 2 show, for a fixed, q=0.63 $\mu m^{-1}$ value and for different temperatures, the R=6 signals (Fig. 2a) and the extra signals relative to R=6.6 (Fig. 2b). Fig. 2a shows that the final decay of the signal is temperature independent ($\frac{d\tau_h}{dT} = 0$) while the $\alpha$ relaxation time, which governs the phonon decay and the subsequent signal raise at short times, strongly increases with decreasing temperature. The extra signals shown on Fig. 2b decay with a relaxation time that increases with decreasing temperature as does $\tau_\alpha$ and we shall discuss in detail in section F the relationship between those two times.

The normal R=6 signals and the R=6.6 extra signals are represented respectively on Figs. 3a and 3b, for different q values at T= 181 K. Fig. 3b shows that the shape of the extra signal is absolutely q independent, contrary to the normal, R=6, signals (Fig. 3a), which exhibit the predicted decrease of the period with increasing q and the $q^{-2}$ dependence of $\tau_h$. The q-independence of the extra signal rules out that it has a diffusive origin. The next Sections will propose a different mechanism which will explain its shape and allow drawing conclusion from its temperature variation.

## D Physical origin of the extra signal

When explaining, in Section B, the physical origin of the ISTS signal, we implicitly supposed that we were dealing with a homogeneous supercooled liquid. This hypothesis may turn out to be incorrect when one considers a solution with a solvent, water, and a solute, LiCl: the liquid may become inhomogeneous under cooling, containing clusters with a solute concentration lower than the original solution embedded in liquid matrix, as suggested in ref. [20]. In the present case, with an increase of the amplitude of the extra signal with R-6, we propose that the homogenous liquid has a higher LiCl concentration than the clusters. Due to this difference in concentration, the environment of a water molecule is different in the two phases, and this reflects in the characteristics of its overtones, thus on the value of the (weak) absorption coefficient of the pump beams by this molecule. As the absorbed energy is instantaneously transferred to the local heat bath, the local temperature of the liquid differs between the two phases. As we show now, this results in an extra time dependence of the "effective" temperature ("effective" density grating) which generates the ISTS signal.

Let us admit, for simplicity that all the clusters have the same salt concentration. The liquid is then composed of two types of regions with different solvent concentrations and thus different absorption coefficients, $k_1$ and $k_2$. The electromagnetic energy density of the two coherent pump beams spatially varies as $(1 + \cos \vec{q}.\vec{r})$, [19, Eq. 2.21b] and, in a usual TG experiment, it is the $\cos \vec{q}.\vec{r}$ component of the absorption of the pump beams which is responsible of the ISTS signal.

In the present case, the change in the local temperature at time t=0 is proportional either to $k_1(1 + \cos \vec{q}.\vec{r})$ or to $k_2(1 + \cos \vec{q}.\vec{r})$ depending on the position $\vec{r}$ of the region (cluster or homogenous liquid) considered. The probe beam propagates in a direction perpendicular to $\vec{q}$, i.e. in a direction for which $\cos \vec{q}.\vec{r}$ is constant and passes through many such two regions whose size are very small with respect to its wave length. The probe beam will thus experience, at time t=0, a mean temperature which also varies along $\vec{q}$ as $\cos \vec{q}.\vec{r}$. The novelty with respect to a usual TG experiment is that this mean local temperature will vary rapidly with time because the thermal energy per unit volume stored in the clusters is different from that of the homogenous liquid so that an equilibration process will take place between them. As the clusters size is very small with respect to $q^{-1}$, this equilibration does not take place as a heat diffusion process but rather as a structural relaxation, with a relaxation time $\tau_a$, of the whole liquid. Consequently, we propose that each individual constituent of the liquid equilibrates its mean thermal energy through interactions with its neighbours, leading to an "effective" short time temperature of the liquid of the form:

$$\tilde{T}(\vec{r},t) = \cos \vec{q}.\vec{r} \left( T_1\delta(t) - \frac{T_a}{\tau_a} \exp\left(-\left(t/\tau_a\right)\right) \right) \qquad (1)$$

while, in a normal ISTS experiment, only exists the first term of Eq. (1). In this equation, the relaxation time $\tau_a$ will turn out to be of the order of $\tau_\alpha$ while $T_a$ is an effective temperature characterizing simultaneously the different temperature increases in the two types of regions and the density of the clusters.

## E Time evolution of the extra signal

It was shown in [16(a)] that, for a LiCl, $6H_2O$ solution, the signal detected in the TG experiment could be deduced from a set of only two hydrodynamics equations coupling the change in local density, $\delta\rho(\vec{r},t)$, in short, $\delta\rho$, with the change in local temperature, $\delta T(\vec{r},t)$, in short $\delta T$. Furthermore, only one of the quantities entering into those two equations, namely the "longitudinal" viscosity, $\eta_L$, needed to have a relaxational behaviour, characterized by the two relaxation times $\tau_\alpha$ and $\tau_\beta$.

More precisely, those two equations read:

$$\overline{\overline{\sigma}} = \left(-c_i^2\delta\rho - \rho_m\beta_m\delta T + \eta_L \otimes \text{div } \vec{v} + \tilde{P}\right)\overline{\overline{I}}, \qquad (2a)$$

$$C_V\delta\dot{T} - T_m\beta_m\delta\dot{\rho} - \lambda\Delta\delta T = C_V\tilde{T}. \qquad (2b)$$

In these two equations, $c_i$ is the isothermal sound velocity, $\beta_m$ is the mean tension coefficient of the liquid, $\rho_m$ and $T_m$ are its mean density and temperature, $C_V$ its specific heat at constant volume per unit volume, and $\lambda$ its heat diffusion coefficient. Furthermore, the symbols $\otimes$ and $\Delta$ stand, respectively, for a time convolution product and for the Laplacian operator while $\tilde{P}$ and $\tilde{T}$ are the source terms, respectively a pressure and energy power: in a usual TG experiment, they act as delta functions in time, with a $\cos \vec{q}.\vec{r}$ space dependence:

$$\tilde{P} = P_1\delta(t)\cos \vec{q}.\vec{r}, \qquad (3a)$$

$$\tilde{T} = T_1\delta(t)\cos \vec{q}.\vec{r}. \qquad (3b)$$

Finally, $\vec{v}$ is the local velocity of the particles, and is related to Eq. 2a through the two equations:

$$\rho_m\dot{\vec{v}} = \text{div}\overline{\overline{\sigma}}, \qquad (4a)$$

$$\rho_m \text{ div } \vec{v} = -\delta\dot{\rho}. \qquad (4b)$$

Performing the Fourier Laplace transform [19] of Eqs. 2a and 2b and solving for a longitudinal wave propagating in the $\vec{q}$ direction leads, with the help of Eq. 4b, to:

$$\sigma = -\left[ c_i^2 \delta\rho + \beta_m \rho_m \delta T + \frac{\omega\eta_L(\omega)}{\rho_m}\delta\rho - P_1 \right], \tag{5a}$$

$$\left(\omega C_V - i\lambda q^2\right)\delta T - \omega\beta_m T_m \delta\rho = C_V T_1. \tag{5b}$$

Eliminating $\delta T$ between those two equations, and further eliminating $\overline{\overline{\sigma}}$ through Eqs. 4a and 4b, one finally obtains, for the ISTS signal ($T_1$ contribution to $\delta\rho$):

$$\delta\rho = iP_L(q,\omega)q^2 \; \beta_m \rho_m \; \frac{\tau_h}{1+i\omega\tau_h} T_1 \tag{6}$$

with:

$$P_L^{-1}(q,\omega) = \omega^2 - q^2\left[ c_i^2 + \frac{\omega\eta_L(\omega)}{\rho_m} + \frac{\beta_m^2 \rho_m T_m}{C_V}\frac{i\omega\tau_h}{1+i\omega\tau_h} \right], \tag{7}$$

where $\delta\rho$ stands here for the Fourier Laplace transform of $\delta\rho(\vec{r},t)$. $P_L(q,\omega)$ is the phonon propagator in the frequency space and, for the frequency range relevant for a TG experiment ($\omega\tau_h \gg 1$), the last term of Eq. 7 reduces to a constant. The only frequency dependent term in the bracket of Eq. 7 is thus simply $\omega\eta_L(\omega)$ and one can replace its two other terms by the square of the adiabatic sound velocity

$$c_a^2 = c_i^2 + \frac{\beta_m^2 \rho_m T_m}{C_V}. \tag{8}$$

$S(t)$, the TG signal, being proportional to $\delta\rho(t)$, its ISTS contribution is proportional to the inverse Fourier Transform of Eq. 6.

We have argued in the preceding Section that, in the case of a LiCl, RH$_2$O TG experiment with R>6, Eq. 3b must be replaced by Eq. 1. Consequently, in Eq. 6, $T_1$ must be replaced by

$$T_1 - T_a\frac{1}{1+i\omega\tau_a}. \tag{9}$$

This leads to the replacement of Eq. 6 by

$$\delta\rho = iP_L(q,\omega)q^2\ \beta_m\rho_m\left[\frac{\tau_h}{1+i\omega\tau_h}T_0 + \frac{\tau_a}{1+i\omega\tau_a}\overline{T}_a\right], \tag{10a}$$

with
$$T_0 = T_1 - \overline{T}_a\ ;\ \overline{T}_a = \frac{T_a}{1-x_a}\ ;\ x_a = \tau_a\big/\tau_h\ . \tag{10b}$$

Eq. 10a shows that the existence of clusters in the solution leads to an ISTS signal which contains two parts. The first one is the usual ISTS signal, convolution product of the phonon propagator with an exponential with relaxation time $\tau_h$ : it has a short time phonon contribution which precedes its long time exponential decay. The second is the convolution product of the same phonon propagator with another exponential with a relaxation time $\tau_a$. Because, in the experiments analysed below, $\tau_a\big/\tau_h$ will never exceed 12%, the second process exists only for times much smaller than $\tau_h$ and never mixes up with the first one. Those two processes are clearly distinguishable on Figs. 1a and 2b, which have been scaled so that the final thermal diffusion process has the same amplitude, for the same wave vector, whatever are the temperature and R values.

Let us note that the analysis of the extra signal, at and below 181 K, will not be possible with Eq. 10a. It will also require the introduction of a second, faster relaxation process, with a relaxation time $\tau_b$ typically ten times shorter than $\tau_a$. More precisely, Eq. 1 will have to be replaced by:

$$\tilde{T}(\vec{r},t) = \cos\vec{q}.\vec{r}\left(T_1\delta(t) - \frac{T_a}{\tau_a}\exp(-\left(t\big/\tau_a\right)) - \frac{T_b}{\tau_b}\exp(-\left(t\big/\tau_b\right))\right), \tag{11}$$

which will lead to

$$\delta\rho = iP_L(q,\omega)q^2\ \beta\rho_m\left[\frac{\tau_h}{1+i\omega\tau_h}T_0 + \frac{\tau_a}{1+i\omega\tau_a}\overline{T}_a + \frac{\tau_b}{1+i\omega\tau_b}\overline{T}_b\right], \tag{12a}$$

with the new relations

$$T_0 = T_1 - \overline{T}_a - \overline{T}_b\ ;\ \overline{T}_b = \frac{T_b}{1-x_b}\ ;\ x_b = \tau_b\big/\tau_h\ . \tag{12b}$$

Eqs. 12a and 12b will be the formula with which we shall fit the extra signals, $T_b$ and $\tau_b$ being equal to zero at the highest temperatures.

## F Quantitative test of the theory and Discussion

### The fit strategy

The total TG signal is the sum of its ISTS and of an ISBS contribution. The latter originates from the $P_1$ term of Eq. 5a and turns out to be simply proportional to $P_L(q,\omega)$: it decays on the $\tau_\alpha$ time scale and is R independent, $P_1$ depending only on the electrostrictive properties of the liquid.

The R=6 signal, i.e. a signal with no extra signal contribution, was analysed in [16(a)] where we showed that Eq. 6, complemented by its ISBS counterpart, fully explained the signal whatever the temperature between 300 and 172 K. The analysis required the use of additional (ultrasonic, Brillouin scattering) experimental results, and also of interpolation methods between well separated temperature domains in order to obtain a good fit of the experimental signals with parameters exhibiting a reasonable thermal variation. To fit the experimental data with reasonably small number of additional parameters, we have assumed that the dependence of the parameters on R is sufficiently weak to be neglected. We have thus separated, whatever are T, R, and $\vec{q}$ in the 181 K case, the R>6 signal into its R=6 contribution and the extra signal. We have first fitted the R=6 contribution with the parameters obtained in [16(a)], which fixes the parameter $T_0$ of Eqs. 10a or 12a, up to a scaling factor. The extra signal then depends, up to the same scaling factor, on the same parameters plus the two extra parameters $\tau_a$ and $\overline{T_a}/T_0$, as well as $\tau_b$ and $\overline{T_b}/T_0$ when Eq. 12a has to be used. Nevertheless, it is the weak R dependence of the $P_L(q,\omega)$ parameters which is responsible of the strong oscillations of the extra signals at short time (1-20 ns). Indeed, the few per cent change in the apparent sound velocity results in a change in the phonon periods between the R=6 and R≠6 signals. Strong, short times, oscillations then remain in the subtraction procedure which leads to the extra signal. Those oscillations prevent the use of a regular least square fit method to determine the additional parameters. Those were obtained through a trial and error method which, in every case, led to a very good fit of the extra signals.

Results and discussion

Table II summarizes the values of the fit parameters for the R=6.6 extra signals, for all the temperatures and $\vec{q}$ values available. The Table also gives the values of $\tau_\alpha$ and $\tau_\beta$ for the same temperatures. Two representative fits of the extra signals are given in Figs. 4 and show the quality of the fits when only one relaxation process (184 K, Fig. 4a) or when two relaxation processes (181 K, Fig. 4b) need to be taken into account. The blue curve in Fig. 4b represents the $\tau_a$ contribution to the 181 K fit. This last figure shows that this contribution fully explains the long-time behaviour of the extra signal but that a faster process is needed to give a good representation of the signal down to 20 ns, at this and lower temperatures. Let us analyse in details the results obtained through those fits.

A first information originates from the value of the ratio $\tau_a/\tau_\alpha$, which is approximately equal to four for the whole temperature range. This finding supports the idea that the structural relaxation correlates with the process which decays with a relaxation time $\tau_a$. Furthermore, it gives us an order of magnitude of the temperature independent size of the clusters. Indeed, in a supercooled liquid, one usually relates $\tau_\alpha$ with the diffusion coefficient, $D$, of the particles through the relation

$$D = \frac{l^2}{\tau} \qquad (13)$$

where $l$ the Fickian length, is basically temperature independent, except possibly [21] in the vicinity of $T_g$. The constant value of 4 for $\tau_a/\tau_\alpha$ suggests that the local temperature equilibrates at $T_0$ when the particles have diffused over a distance approximately equal to $2\,l$ and this can be interpreted as an approximate measure of the size of the clusters. We have estimated the value of $l$, $l \approx 1.2$ nm, using NSE measurements performed by Mamontov *et al.* [22] on a R=7.3 solution. The details of our calculation are given in the Appendix.

An interpretation of $\tau_a$ slightly different from the one used above consists in proposing that this time represents the life time of the clusters, the diffusion of the particles, inside the clusters, and in their environment, destroying the clusters after that time. This second interpretation does not change either the estimate of their size, or the fact that this size does not change with temperature.

- Secondly, the scaling factor, $R_0$, between the R=6.66 and the R=7.14 extra signals, last column of Table II, increases only very slowly from 2.0 to 2.6 when the temperature decreases from 190 to 172 K. This scaling factor is just the number by which one multiplies the R=6.66 extra signal, for times larger than 20 ns, to obtain the R=7.14 extra signal. Its very existence implies that $\tau_a$ and $\tau_b$ do not change from one signal to the other while, c.f. Eq. 12b, the ratios $\overline{T}_a/T_0$ and $\overline{T}_b/T_0$, which are both positive, are multiplied by the same factor. These positive signs imply that, after the instantaneous heating of the liquid by the pumps, the clusters have a mean temperature lower than the other part of the liquid and that an energy flow has to take place from the homogenous liquid to the clusters to obtain an homogenous mean temperature, $T_0$ (we delay the discussion of the physical meaning of $\tau_b$ to the end of this subsection). Also, this flow is governed by $\tau_a$ and $\tau_b$: their identical values for R=6.66 and R=7.14 imply that the size and the composition of the clusters are the same for the two concentrations and that it is only their number which increases with increasing R. The quasi independence of the scaling factor on the temperature indicates that this increase is practically the same at all temperature.

- Thirdly, we see on the same Table that the ratio $\left(\overline{T}_a + \overline{T}_b\right)/T_0$ is constant between 190 and 181 K, decreases slightly at 178 K and more at 172 K. This ratio measures the relative energy transfer from the homogenous part of the liquid to the clusters, and thus monitors the number of clusters in the liquid. Its independence on the temperature for the three highest ones indicates that this number is approximately constant at high temperature, and starts decreasing below.

The picture which emerges from this analysis is thus that the clusters do not vary in size, composition and numbers for a given value of R, down at least to 181 K, changes starting to appear below, and that, whatever the temperature, increasing R simply increases the number of clusters without changing their characteristics.

Let us finally discuss the second relaxation process characterised by $T_b$ and $\tau_b$. There is no correlation between $\tau_\beta$ and $\tau_b$: either the b process does not exist (190 and 184 K), or $\tau_b$ is not proportional to $\tau_\beta$, their ratio varying from $\approx 3$ (181 K) to $\approx 35$ (172 K). The absence of a "b process" at high temperature and the lack of proportionality between $\tau_\beta$ and $\tau_b$ suggest that it could be a sort of boundary effect. The molecules located at the boundary between a cluster, with their own temperature, and their surrounding with another one, could have their local temperature relaxing more rapidly than those located inside the cluster. This effect would not take place at high enough temperature where the molecules relax so fast that the difference between those boundary and the inside molecules would not appear.

<u>Comparison with other results</u>

As already mentioned in the Introduction, the study of LiCl, $RH_2O$ solutions has a very long history. Following the publication of the revised equilibrium LiCl-$RH_2O$ phase diagram by Moran [23] and the first experiments recalled in the Introduction, many studies of this supercooled phase were devoted to their macroscopic properties such as their thermal expansion [24], low frequency dynamics [16(a), 25], dielectric broad band spectrum [26]. Some of them [16(a), 22, 25] demonstrated the existence, on top of the usual $\alpha$ relaxation process, of the $\beta$ relaxation process, already detected by other techniques, such as QENS [27], NMR and neutron spectroscopy [28], by the Université de Lyon group. Another series of publications were concerned with, or driven by, a different point of view, more closely related to the present paper: does a phase transition, involving the structure of the solution or of the solvent, i.e. of water, take place when cooling this solution? Some studies preceded [9-b, and references therein] the prediction of a water-water liquid transition in pure water in the no-man's land region, [4], while others [26, 29] were driven by such a consideration. The latest experiment [26] presently concludes to the absence of such a transition, a point that we shall discuss again in the last part of this Section.

Other studies focused on a more microscopic aspect of the structure of those solutions, through the study of the environment of a given ion ($Li^+$ and/or $Cl^-$), or atom belonging to the water molecule (H and/or O). Neutron [30-32], and X Rays [33] scattering measurements revealed not only the local environment of some constituents [30] but also the radial distribution functions for pair of those constituents [31-33]. Some experiments were performed at discrete values of R (e. g. R=4 or R=6). Other techniques such as NMR [28], Infrared spectroscopy [34] or quantum mechanical molecular dynamics computations [35], concentrated on the environment of a specific constituent; in that case, more or less implicitly, the authors considered this constituent plus its immediate environment as one component, the rest of the liquid being taken as a homogenous liquid. For instance, in ref. [34], the authors studied one IR band of the water molecule. They divided the solution into one LiCl entity plus the water molecules contained in its first hydration shell as one entity, the rest of the liquid being considered as pure water. Similarly, in ref. [35], the quantum mechanical computation considered one single $Li^+$ ion immersed in a sea of water molecules.

Nevertheless, none of the studies mentioned so far was concerned with a situation similar to the present one where the liquid is no longer a single, homogenous, solution but a liquid in which coexist two phases with different solute concentrations. The situation is different concerning glasses obtained by a rapid quench of a LiCl, $RH_2O$ solution. R>9 [36] and later 9≥R≥3 [37] solutions have been studied. Similarly to ref. [34], the 2500-3600 $cm^{-1}$ OH stretching band of the water molecules was studied by Raman spectroscopy. Ref. [36] proposed a phase separation in the glass between a pure water-like LDA phase and an R=8, homogenous, amorphous phase. The more careful (and performed ten years later) study of ref. [37] now proposes the replacement of the R=8 homogenous phase by a phase in which the first hydration shell would be formed of water molecules in a given HDA form while the other water molecules would be in another HDA one, called "HDA under low pressure". A conclusion partly similar has been recently reached in the molecular dynamics calculation of Le and Molinero [20]. These authors made use of a coarse grained potential both for the water molecules and for the solute, considered as formed of a unique type of particles, the potentials of both components containing no long range interaction. This allowed Le *et al.* to perform computations in the glass phase after a rapid quench, with more than $10^5$ particles on a time scale of a few μs. They obtained a phase diagram in which for R ≤9, the hypothetical glass would split into two types of clusters, particle (ion) rich ones and LDL ones, the size of these clusters being of the order of a few nanometres. This last result seems to be the first numerical suggestion of a phase separation rather similar to the one we have detected, though it takes place in the glass phase and not in the liquid one. The possible existence of clusters in a R=7.3 solution has also been recently suggested by Mamontov e*t al*. [22]. These authors noticed, below approximately 210 K, the presence, in their QENS experiments performed on a NSE instrument, of an elastic component for wavevectors Q larger than 0.2 $\text{Å}^{-1}$.They deduced from this finding that dynamical heterogeneities could exist below that temperature with an estimated size of $\approx 3$ nm, a value rather similar to the one we are proposing. They nevertheless made no suggestion on the relative composition of these dynamical heterogeneities and of the rest of the solution.

Also, the Le and Molinero method cannot discriminate between the different ions present in the solution. As indicated in the Introduction, the extra signal exists for LiCl, $RH_2O$, at and below 190 K, [38], at least for $6.45 \leq R \leq 7.14$, while it is not detected in LiBr, $RH_2O$ for R=7.14, the concentration at which the extra signal is most intense in the $Cl^-$ solutions. This striking difference may be simply due to the detection mechanism: the latter is the difference between the energy absorbed from the laser beam by the clusters and by the rest of the liquid. It is possible that the modification of the overtones, which is responsible for this difference, is much less important in the bromine solution than in the chlorine one. It is nevertheless more reasonable to propose a different, and possibly related, explanation based on the observation of the same difference in the depolarized Raman spectrum of these two solutions. The chlorine spectra were measured as a function of the temperature in [26] at R=7.3, i.e. at a concentration very close to the R=7.14 studied here. The results can be compared with those obtained with LiBr, $6H_2O$ [16 (b)], at temperatures between 230 and 250 K. The spectra are, in part, similar for the two salts, with a minimum around 200 GHz in both cases and a microscopic peak, located at 5 THz for the chlorine solution and 12 THz for the bromine one. A striking difference is nevertheless found in the intensity of the two Boson peaks. In the chlorine case, it has the same intensity as the microscopic one and a frequency of the order of 1 THz. Conversely, this peak is totally absent in the bromine spectrum. The Boson peak has been generally associated to the existence of a local cooperativity in the motion of the constituents of the super cooled liquid, or of the glass. A correlation between the existence of this peak and the extra signal is quite tempting: it would be in line with the great ability of a chlorine ion to replace a water molecule, already pointed out in the Introduction, while the larger size of the bromine ion may impede such a correlation with the water molecules motion.

**G Summary and Final Remarks**

We have identified in supercooled LiCl, $RH_2O$ solutions with R=6.45, 6.66, and 7.14, the existence of clusters, with a nanometre size and a composition different from the mean solution composition, such clusters being absent in the R=6 solution. These clusters were identified through a Transient Grating experiment which was interpreted as the result of a different absorption of the pump beams by the clusters and by the mean liquid. The signals then contain a new signature, characteristic of the later temperature equilibration between those clusters and the mean liquid. This equilibration takes place on a time scale $\tau_a$ approximately four times longer than the α relaxation time, $\tau_\alpha$, of the supercooled solution. The clusters show up for temperatures between 190 and 172 K, lowest temperature of our experiments. We find that they have the same composition and size whatever the temperature or the value of R, their number simply increasing with R-6.

It is the first time that a clear observation of such a clustering effect has been observed in a supercooled water solution. It has not been found, for instance in the corresponding LiBr, $RH_2O$ solutions. Our result is also at variance with the case of supercooled glycerol, $RH_2O$ solutions with 4.2≤R≤6.4 [39], for which Murata and Tanaka did not report a clustering effect but a Liquid-Liquid transition. Conversely, evidences for such clusters have been found for the same LiCl, $RH_2O$ solutions when rapidly quenched to the glass phase and it has been proposed that these clusters are formed of pure water, but the structure of this amorphous ice is still highly controversial. A recent experiment has suggested a specific form of HDA [37], while a computer simulation [20] proposes rather a LDA structure.

It is thus important to better characterize the clusters we have found in the superccooled phase. Small angle neutrons with isotope substitution would be well suited for the study of the local structure and of the size of the clusters, that we propose to be of the order of a few $l$, where $l$ is the Fickian length associated with the translational length of the water molecules. Also Raman spectroscopy measurements on the OH stretching band already used in [37] to detect inhomogeneities in some LiCl, $RH_2O$ glasses could allow for a comparison with the glass case. Some of those experiments are presently planned

**Acknowledgments**

The Transient Grating experiments have been performed in Firenze (I) under the LENS contract 1459. During all the experiments, a very important help was given to us by P. Bartolini, R. Cucini and A. Taschin that we want to warmly thank here. This work was financially supported by the French Agence Nationale de la Recherche (ANR JCJC0135).

**Appendix**

The approximate value of l given in the main text has been obtained with the help of ref. [22]. Mamontov *and al.* have measured a translational diffusion length in a R=7.3 solution by two different techniques, Nuclear Magnetic Resonance (NMR), and Neutron Spin Echo (NSE) with a wave vector Q= 0.1 Å$^{-1.}$ Both sets of data agree, within experimental errors, between 300 and 200 K. Following a suggestion made, e.g.; in [27], these data can be nicely fitted by a Vogel Fulcher law

$$D = D^0 \exp\left( -\frac{A}{T - T^0} \right) \qquad \text{A-1}$$

with $D^0$ =1.35 $10^{-7}$ m$^2$ s$^{-1}$, $A$ =400 K, $T^0$ = 110 K.

The Stokes Einstein relation predicts that the product $D\tau$, where $\tau$ is a relaxation time, is independent of temperature:

$$D\tau = cte \qquad \text{A-2}$$

Nevertheless, the preceding statement is imprecise in the sense that, in a supercooled liquid there exists many different relaxation times, each related to one of the many possible observables. These relaxation times may differ by order of magnitudes, ones from the others, depending on the observables, though the ratio of two of them is basically temperature independent. In order that the constant of the r. h. s. of Eq. (A-2) represents the square of a Fickian diffusion length, it is necessary that both quantities are measured by the same technique, here NSE, with the same Q. Those relaxation time values are also given in [22], and we have verified that, in the 300-200 K range of interest here, they differ from the values obtained by the HD-TG technique of [16 (a)] by a factor 500. The Fickian length reported in the main text has thus been obtained with the help of the expression:

$$l = \left[500 D_{NSE} \, \tau_\alpha\right]^{0.5} \qquad\qquad \text{A-3}$$

where, for each temperature, $D_{NSE}$ is the value obtained in [22] while $\tau_\alpha$ is taken from Table II. We have further verified that in the 190-172 K range of interest here, the $l$ value was constant within a factor 1.6, compatible with a slight violation of the Stokes Einstein violation [21] at the lowest temperatures (172 K$\approx$1.23 $T_g$). $l_{T=190K}$ represents then a reasonable estimate of the Fickian length in our experiment and Eq. (A-3) yields

$$l = 1.2 \text{ nm.}$$

<u>Figures Captions</u>

Fig. 1

 a) The different TG signals at T=184 K, for q= 0.63 μm$^{-1}$: R=6 —, 6.45 —, 6.66 —, and 7.14 —.

 b) The three extra signals deduced from Fig. 1a. [(R=6.45)-(R=6.00)] extra signal, —; [(R=6.45)-(R=6.00)] extra signal, —,rescaled for t=20 ns, with the [(R=7.14)-(R=6.00)] one; [(R=6.66)-(R=6.00)] extra signal, —; this [(R=6.66)-(R=6.00)] extra signal rescaled in the same way as above, — totally coincides with the [(R=7.14)-(R=6.00)] one, —.

Fig. 2

 a) The R=6.00 TG signals at five different temperatures, T=190K —, 184 K —, 181 K —, 178 K —, and 172 K —, for q= 0.63 μm$^{-1}$.

 b) The [(R=6.66)-(R=6.00)] extra signals for the same wavevector and temperatures, and with the same symbols as on Fig. 2a.

Fig. 3

 a) The R=6.00 TG signals at T=181 K for four different wavevectors: 0.63 —, 1.0 —, 1.38 —, and 1.76 —.μm$^{-1}$

 b) The [(R=6.66)-(R=6.00)] extra signals for the same temperature and wavevectors, and with the same symbols as on Fig. 3a. The extra signals have been shifted, for clarity, by 0.02 a. u. in order to show their identical behavior for t>20 ns.

Fig. 4

 a) Fit of the [(R=6.66)-(R=6.00)] extra signal, at T=184 K, for q= 0.63 μm$^{-1}$, by one exponential (see Eq. 1); —, extra signal, —, fit one exponential.

 b) Fit of the [(R=6.66)-(R=6.00)] extra signal, at T=181 K, for q= 0.63 μm$^{-1}$, by two exponentials (see Eq. 11), same symbols as in Fig. 4 a. The — and — lines represent the contributions, respectively, of the first and second exponentials with their respective weights.

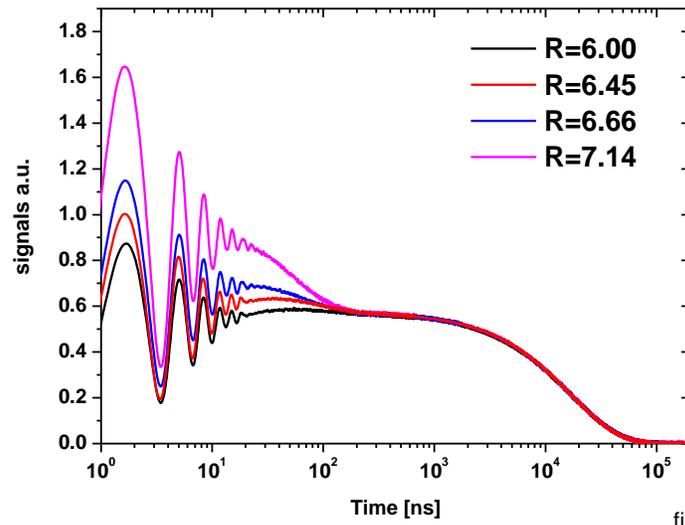

fig.1a

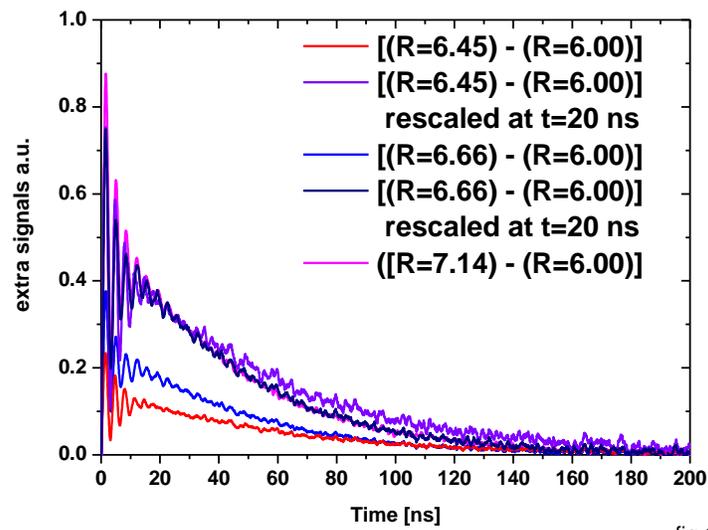

fig 1b

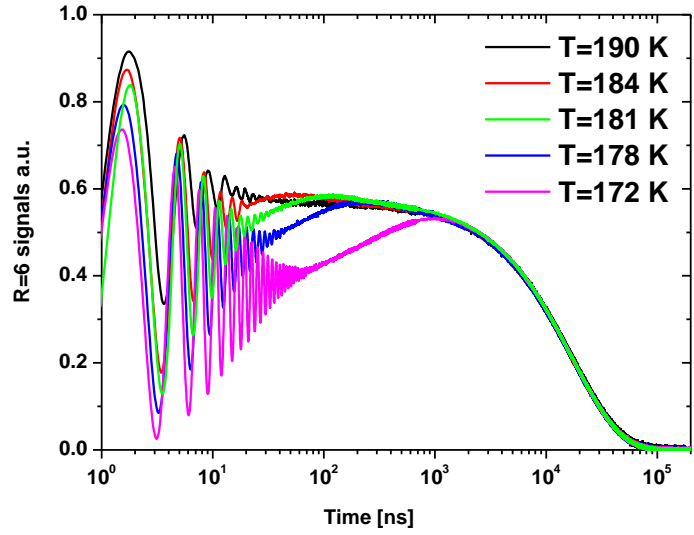

fig 2a

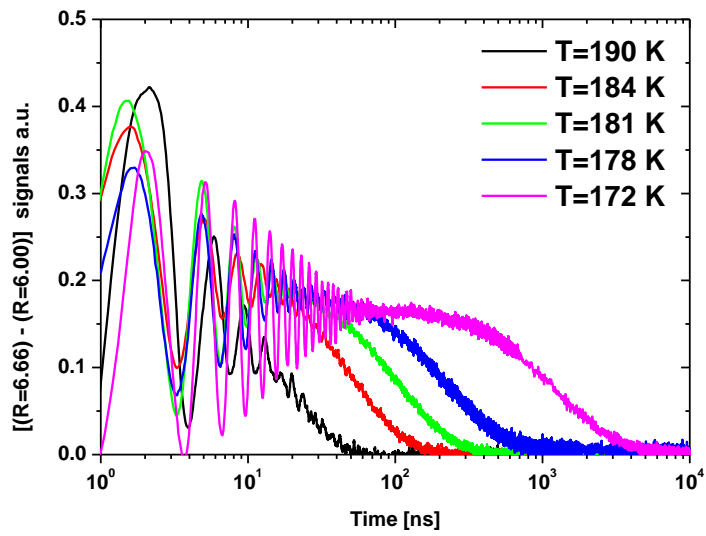

fig 2b

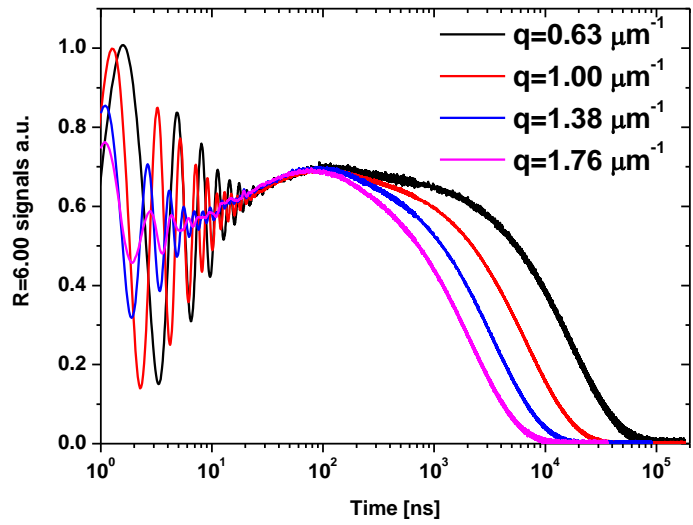

fig 3a

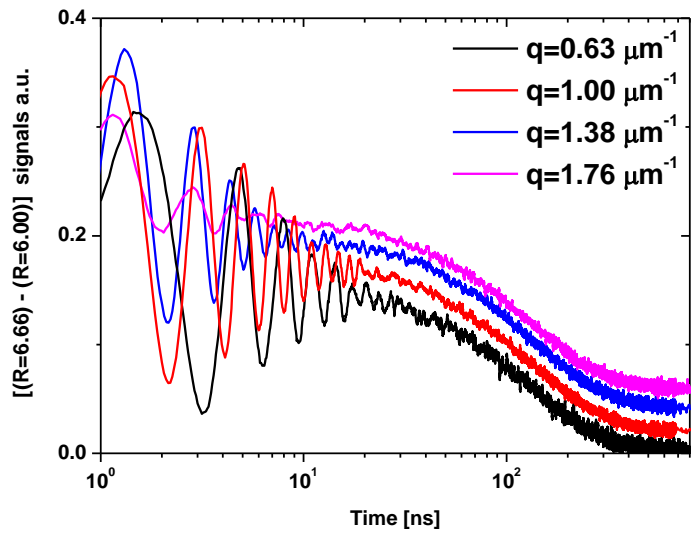

fig 3b

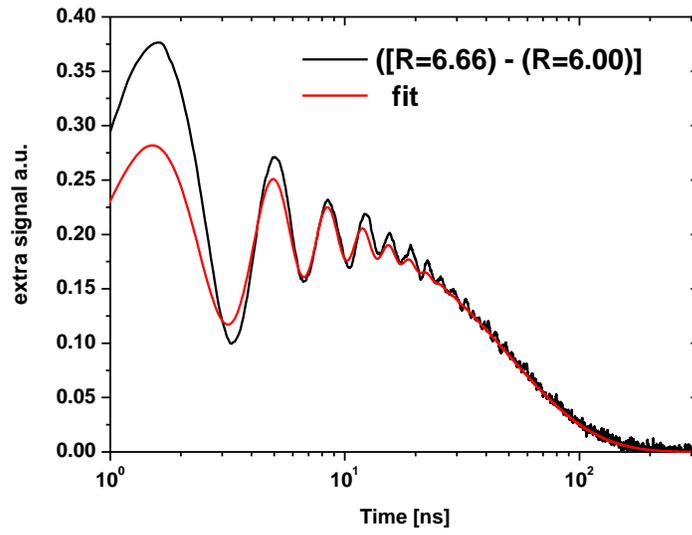

fig 4a

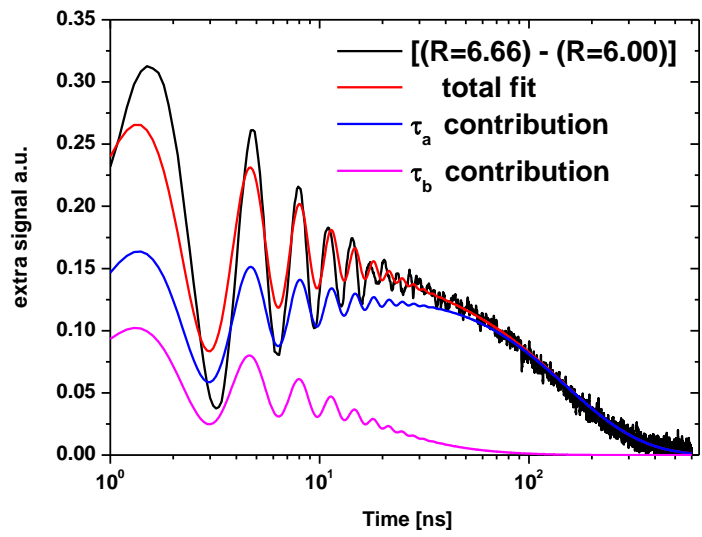

fig 4b

<u>Tables Caption</u>

Table I : List of the extra signals analyzed in the present paper

Table II : Parameters $\tau_a$, $\overline{T}_a / T_0$, $\tau_b$, and $\overline{\overline{T}}_b / T_0$, obtained in the fitting procedure of the R=6.66 extra signals, and ratio, $R_0$, between the R=7.14 and the R=6.66 extra signals amplitudes. The parameters $\tau_\alpha$ and $\tau_\beta$ have been obtained in ref. [16 (a)].

**Table I**

| T (K) | R=6.00 | R=6.45 | R=6.66 | R=7.14 | |
|---|---|---|---|---|---|
| **190** | X | X | X | | |
| **184** | X | X | X | X | |
| **181[a]** | X | X | X | X | q=0.63 μm⁻¹ |
| **178** | X | X | X | X | |
| **172** | X | X | X | X | |
| **q (μm⁻¹)** | **0.63** | **1.00** | **1.38** | **1.76** | |
| **R=6.00** | X | X | X | X | T=181 K |
| **R=6.66** | X | X | X | X | |

**a** This series of experiments was performed, in fact, at an unknown temperature slightly above 181 K. We could only exploit the ratio of the extra signals between R=6.66 and R=7.14 but not the extra signals themselves.

**Table II**

| T (K) | $\tau_\alpha$ (ns) | $\tau_a$ (ns) | $\dfrac{\overline{T}_a}{T_0}$ | $\tau_\beta$ (ns) | $\tau_b$(ns) | $\dfrac{\overline{T}_b}{T_0}$ | $\dfrac{\overline{T}_a+\overline{T}_b}{T_0}$ | $R_0$ |
|---|---|---|---|---|---|---|---|---|
| **190** | 3.64 | 16 | 0.57 | 1.65 | | | 0.57 | |
| **184** | 21.6 | 38 | 0.55 | 2.6 | | | 0.55 | 2.0 |
| **181** | 32.5 | 120 | 0.35 | 3.3 | 12 | 0.22 | 0.57 | 2.1 |
| **178** | 58 | 220 | 0.31 | 3.4 | 22 | 0.15 | 0.46 | 2.4 |
| **172** | 360 | 1260 | 0.15 | 11.4 | 400 | 017 | 0.32 | 2.6 |